\documentclass[article]{article} 

\usepackage{mathrsfs} 
\usepackage{graphicx} 
\usepackage{mediabb}
\usepackage[top=3cm, bottom=3cm, left=2cm, right=2cm]{geometry}
  
\def\be{\begin{equation}} \def\ee{\end{equation}} 
\def\bc{\begin{center}} \def\ec{\end{center}}  \def\ef{\end{figure}}   
\def\Msun{ M_{\odot \hskip-5.2pt \bullet} }  
\def\Msun{M_\odot}

   \def\deg{^\circ}

 \def\V0{ V_0 }

 \def\kms{ km s$^{-1}$ }    %\def\/{\over}
  
\def\RE{ R_{\rm E} } \def\rhoE{ \rho_{\rm E} } \def\ME{M_{\rm E} } 
\title{Gravitational Focusing of Low-Velocity Dark Matter on the Earth's Surface\footnote{To appear in Galaxies, special issue on 'Debate on dark matter'.}} 
 
\author{Yoshiaki Sofue \\  
Institute of Astronomy, The University of Tokyo, Mitaka, Tokyo 181-0015, Japan\\ E-mail: sofue@ioa.s.u-tokyo.ac.jp}  

\begin{document} %%%%%%%%%%%%%%%%%%%%%%%%%%%%%%%%%%%
\maketitle

\abstract{We show that the Earth acts as a high-efficiency gravitational collector of low-velocity flow of dark matter (DM). The focal point appears on the Earth's surface, when the DM flow speed is about 17 km/s with respect to the geo-center. We discuss diurnal modulation of the local DM density influenced by the Earth's gravity. We also touch upon similar effects on galactic and solar system objects.}


%\begin{document} %%%%%%%%%%%%%%%%%%%%%%%%%%%%%%%%%%%
\section{Introduction}
                      
The local dark matter (DM) density in the Solar vicinity has been extensively estimated in the last decades using the rotation curve of the Milky Way, and has a converging value of $\rho_\odot~\sim~0.4$~GeV~cm$^{-3}$~~\cite{Sofue2017,Sofue2020,Salucci2019}. 
However, thus measured density gives only an averaged value around the Solar system integrated over the phase space. Information of more specific behaviors of DM as a function of the space (direction) and velocity would be crucial for the direct detection experiments as well as for determining the physical properties of the detected particles from the event count rates~\cite{Alenazi+2008,Graham+2016,Bandyopadhyay+2012}. 
The DM density distribution as a function of the particle's velocity, or the spectral DM density (SDD) as well as its time variation (modulation) in the laboratory frame on the Earth's surface would provide useful information for such experiments.  

Annual, monthly, and daily modulations of the flux of SDD due to the motion of the Earth as well as to the gravitational perturbation in the Solar system have been thoroughly investigated
~\cite{Alenazi+2006,Peter2009a,Peter2009b,Peter2009c,Bandyopadhyay+2012,Lee+2014}. 
It has been suggested that the daily modulation of the gravitational focusing by the Earth leads to amplification of the local density to a detectable level~\cite{Kouvaris+2016}. 
It has been also shown that gravitational focusing of WIMP (weakly interacting massive particles) with zero velocity dispersion in the rest frame enhances the density by $\sim 10^8$ times at a certain focal point in the interplanetary space~\cite{prezeau2015}. 
Besides the gravity, scattering by the Earth's nucleons is also suggested to cause modulations~\cite{Kavanagh+2017}.
We~here, however, consider only the gravitational interaction.  
{If the DM has a directional streaming, the differential SDD at a certain velocity (range) is amplified by the focusing effect due to the gravitational force by a celestial object.} 

We here consider the gravity of the Earth, which has the largest solid angle among the celestial objects, assuming that the detector is located on the Earth's surface.
 We show that DM flux at a particular geo-centric velocity converges on the Earth's surface with high amplification. 
 Such focusing effect would be useful to be considered not only for the detection purpose but also for interpreting the physics of the measured flux.

\section{Focusing by the Earth}
 
 We consider a case that the DM streaming is monochromatic, and trace the motion of DM particles by solving the equations of motion
\begin{equation}
\frac{d^2{\bf r}}{dt^2}=-G \frac{M(r)}{r^2}\frac{{\bf r}}{r} \ ,
\label{eqmotion}
\end{equation}
where ${\rm r}=(x,y,z)$ is the geo-centric coordinates.
$M(r)=\int_0^r 4 \pi p^2 \rho(s)ds$ is the mass of the gravitating body, which is assumed to have spherical density distribution. 
We neglect the Coriolis force and parallactic aberration due to the spin of the Earth.
 
\subsection{Point Mass}
\begin{figure} 
\begin{center} 
\includegraphics[width=12cm]{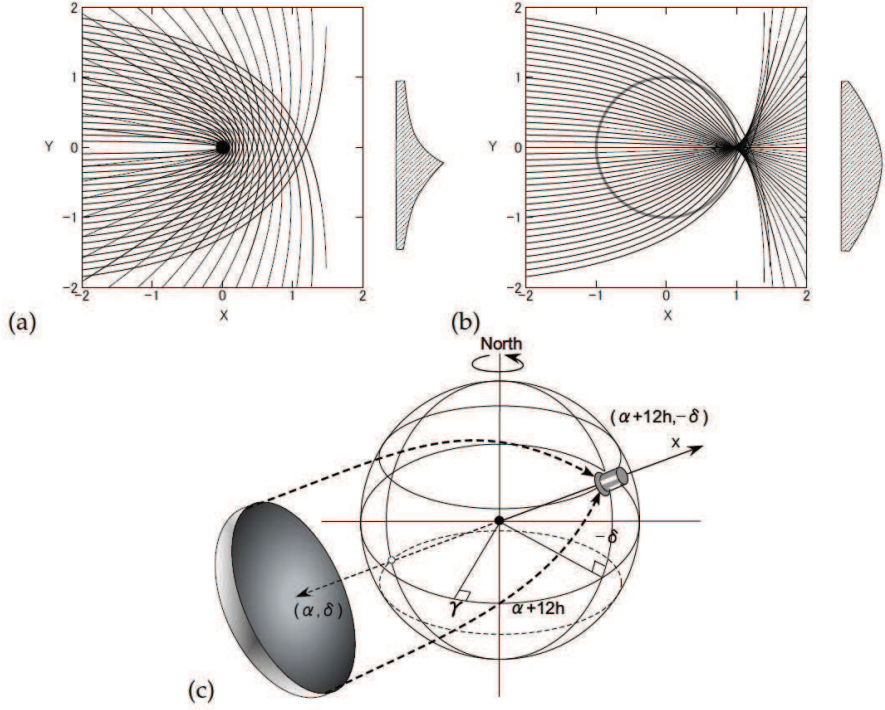} 
%(a)\includegraphics[width=5cm]{orbitPoint.pdf} 
%\includegraphics[width=0.7cm]{concave.pdf} 
%(b)\includegraphics[width=5cm]{orbUni.pdf}   
% \includegraphics[width=0.7cm]{convex.pdf} \\
%(c)\includegraphics[width=7cm,angle=0]{telescope.pdf}
\end{center} 
\caption{(\textbf{a}) Gravitational focusing of dark matter (DM) stream by a point mass such as a star in the interstellar space, and (\textbf{b}) a nearby large-solid angle body like the Earth with constant density (top right). Point mass acts as radially-concave and azimuthally-convex collector, resulting in diverging orbits. On the other hand, a uniform mass sphere acts as a convex collector, having a focal point. (\textbf{c}) Schematic illustration to show that a finite-density massive ball acts as a convex lens having a focal point.}
   \label{orbPandU}
   \end{figure}

We first examine a case of a point mass, or $M(R)=const.$ as applied to the gravitational deflection of DM flux by the Sun~\cite{Alenazi+2006}. A point mass acts as a semi-concave collector, and the DM particles are scattered because of the decreasing deflection angle with increasing impact parameter, resulting in diverging orbits (Figure \ref{orbPandU}). Hence, focusing occurs only of DM particles with a particular impact parameter, which results in focusing from a small circle (ring on the sky) around the gravitating body.

\subsection{Uniform Density Sphere}
On the other hand, an object like the Earth with sufficiently large solid angle with uniform density acts as a real convex collector having a focal point. 
The orbits of DM particles are deflected by the Earth's gravity when the geo-centric velocity is slow enough around $v\sim \RE \sqrt{2 \pi G\rhoE}\sim 10\ {\rm km \ s^{-1}}$. Here, $\RE=(R_{\rm eq.}\times R_{\rm pole})\simeq(6378\times 6356)^{1/2}=6367$ km is the mean radius and $\rhoE\sim 5.53$ g cm$^{-3}$ is the mean density for a total mass of $\ME=5.974\times 10^{27}$ g. 

Figure \ref{orbPandU}a,b show the orbits of DM particles with initial injection velocity of 6 \kms in a spherical gravitational potential in two extreme cases, where the Earth is assumed to be a point mass of $\ME$ and a sphere of uniform density $\rhoE$, respectively.

\subsection{Gravitational Focusing by the Earth}
\begin{figure} 
\begin{center}   
\includegraphics[width=8cm]{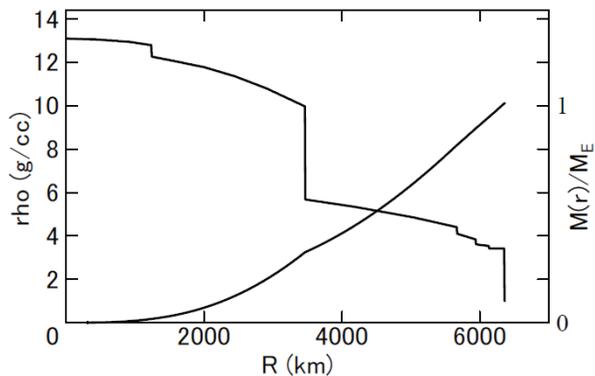} 
\end{center}
\caption{Density profile in the Earth~\cite{Kennett1998} and enclosed mass in a sphere of radius $R$.}
   \label{rhoEarth}
   \end{figure}

More realistic orbits of DM particles can be traced by solving the equations of motion in gravitational field of the real Earth. We assume that the Earth is a sphere of radius $\RE$ with density profile as shown in Figure \ref{rhoEarth}
~\cite{Kennett1998}.  %(Kennet 1998). 
%\hl{We}
%\textcolor{red}{Reference [14] is  missing. Please renumber the references so they appear in sequential numerical order.} 
We take the Cartesian coordinates $(x,y,z)$ with the geo-center being at (0,0,0), $x$ axis is through the detector on the Earth surface toward the zenith, and $y$ and $z$ are perpendicular axes.   %MDPI:  %Reference [14] is  missing. Please renumber the references so they appear in sequential numerical order.
.

\begin{figure} 
\begin{center}   
 \includegraphics[width=12cm]{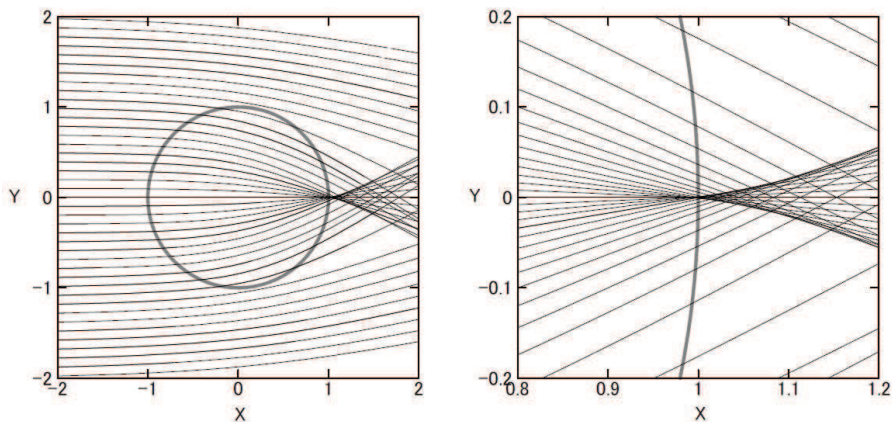}     
\end{center}
\caption{Geo-gravitational focusing of DM stream by the Earth with the density profile given in Figure~\ref{rhoEarth} for $v=v_x= 17$ \kms on the Earth's surface with aperture diameter of $\sim$6000 km. The~right panel is close up near the Earth's surface.} 
   \label{focus}
   \end{figure}

 We trace the orbits of DM particles put on a plane sufficiently remote from the Earth, or at $x=x_{\rm ini}=-100\RE$, by integrating the differential equations \ref{eqmotion} with initial velocities $v_x=v_{\rm inj}$, $v_y=0$ and $v_z=0$. 
Calculations were obtained for various values of $v_{\rm inj}$.
An efficient focusing on the Earth's surface was found to occur, when the initial velocity is close to $v_{\rm inj}=17$ \kms. 
Figure \ref{focus} shows the orbits of DM in such a special case. 

The figure shows that DM particles sharply focus on the Earth's surface, where they cross each other, exhibiting high-efficiency conversion and amplification of the flux.
Thus, the Earth plays a role of an ideal collector of DM flow at 17 \kms.  It may be emphasized that the DM flow within $0.2 \RE$, or aperture $D\sim 0.4 \RE\sim 5000$ \kms, the collector is almost aberration free, having a sharp focal point (Figure \ref{focus}). If some aberration is allowed, the aperture diameter is as large $D\sim \RE\sim$ 12,000 km. 

\section{Spectral Density and Flux Amplification}

\subsection{Spectral DM Density}

{We assume that the DM particles are distributed uniformly in the space around the Earth, but their velocities are not uniform and anisotropic, obeying the Maxwellian distribution represented 
by
\be
%f({\bf v})=\frac{1}{\sqrt{\pi}}
\rho({\bf v})=\frac{\rho_\odot}{(2 \pi \sigma_{\rm DH}^2)^{3/2}}
\exp 
\left(
-\frac{\left[(v_x+v_{x{\rm E}})^2+(v_y+v_{y{\rm E}})^2+(v_z+v_{z{\rm E}})^2 \right]}{2\sigma_{\rm DH}^2} 
\right).
    \label{fv}
\ee 

Here, ${\bf v}=(v_x,v_y,v_z)$ is the 3D velocity of a DM particle 
with respect to the geo-center (\mbox{$v_x\sim 17$ \kms}, $v_y \sim v_z \sim 0$ within $\delta v$), ${\bf v}_{\rm E}=(v_{x{\rm E}},v_{y{\rm E}}v_{z{\rm E}})$ is the 3D velocity of the Earth with respect to the dark halo in the rest frame fixed to the Milky Way, which include the solar motion, galactic rotation of the local standard of rest (LSR), and Earth's rotation around the Sun, and $\sigma_{\rm DH}$ is the velocity dispersion of the dark halo.
}

\begin{figure} 
\begin{center}  
\includegraphics[width=9cm]{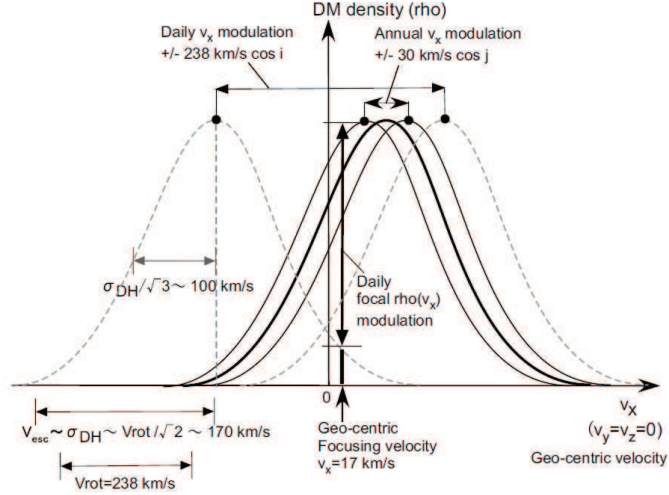} 
\end{center}
\caption{Schematic illustration of local modulation of the flux of low-speed DM caused by the varying injection speed $v_x$ along the $x$ axis fixed to the direction from geo-center to the detector on the Earth's~surface. }     
\label{modu}
   \end{figure}

According to the rotation of the coordinates due to the Earth's spin, injecting DM flux varies with time and position of the detector. The daily velocity variation (modulation) is large, following the change in the $v_x$ axis direction between the up- and down-stream directions of the Galactic rotation at $V_{\rm rot}=238$ \kms in the DM halo at rest. The amplitude attains maximum, when the $v_x$ axis lies in the plane including the Sun's motion toward $(l,b)\sim(90^\circ,0^\circ)$ for a detector located at geographic latitude $\beta \sim \pm 48^\circ$ (Figure \ref{modu}). The directional aberration of the DM flux due to the spin of the Earth is about $\sim$$0.46~{\rm km \ s^{-1}}/v \ {\rm cos}\  \beta  $ toward the east, where $\beta$ is the geographic latitude, and may be neglected compared with the detector's angular resolution.

\subsection{Amplification}

We introduce an amplification factor $\mathcal{A}(v)$ of DM flux defined by 
\begin{equation}
    \mathcal{A}(v)=\frac{S_{\rm aperture}}{S_{\rm fucus}}
    =\left(\frac{r_{\rm aperture}}{r_{\rm focus}}\right)^2,
    \label{eqamp}
\end{equation}
where $S_{\rm aperture}=\pi r_{\rm aperture}^2$ is the collecting area (aperture) of the injection flux enclosed by a circle of the impact parameter $r_{\rm ini}=r_{\rm aperture}$, and $S_{\rm focus}=\pi r_{\rm focus}^2$ is the focal area, respectively.

\begin{figure}  
\begin{center}    
 \includegraphics[width=8cm]{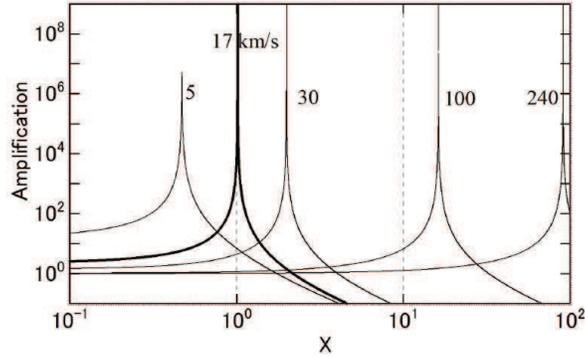}   
\end{center}
\caption{
DM flux amplification by the Earth for injection velocities 5, 17, 30, 100, and 240 \kms. DM flow with $v_{\rm inj}=17$ \kms focuses on the Earth's surface with the focal length equal to $R_{\rm E}$ at high amplification. Lower/higher velocity flows focus inside/outside the Earth with shorter/longer focal length than $R_{\rm E}$.} 
   \label{flux}
   \end{figure}

In Figure \ref{flux} we plot the amplification factor $\mathcal{A}(v)$ as a function of the distance from the geo-center for DM flows with injection velocity $v=5$, 17, 30, 100 and 240 \kms for near axis orbits with initial $y=0.1 R_{\rm E}$. The focal position moves with the injection velocity, and the maximum amplification at the focus on the Earth's surface is obtained for a particular injection velocity of \mbox{$v_{\rm inj}=17$ \kms.} The~amplitude reaches as high as $\mathcal{A}\sim 10^9$ times the initial flux in the present computation. In~principle, however, the amplification reaches infinity at the focus according to the geometry of particles' trajectories crossing the $x$ axis.

{By the focusing, DM particles enclosed by an aperture of radius $r_{\rm aperture}$ in a velocity range $v_x=v-\delta v/2$ and $v+\delta v/2$ are collected, attaining the maximum amplification at the focus to yield DM density as high as
\begin{equation}
\Delta \rho_{\rm aperture} 
\sim 
\left( \frac{r_{\rm aperture}}{r_{\rm focus}} \right)^2 \Delta \rho
\sim
\left( \frac{r_{\rm aperture}}{r_{\rm focus}} \right)^2
\int_{v-\delta v/2}^{v+\delta v/2}
\int_{\delta v/2}^{\delta v/2}
\int_{\delta v/2}^{\delta v/2}
\rho({\rm v}) dv_x dv_y dv_z.
 \label{amp}
\end{equation} 

The integral part is on the order of 
$\mathcal{O}(1) \rho_\odot (\delta v/\sigma_{\rm DH})^3$
for small $\delta v_x$ etc., so that the equation may be rewritten for an order-magnitude estimation as
\be
\Delta \rho _{\rm Earth}
\sim \rho_\odot 
\left(\frac{r_{\rm aperture}}{r_{\rm focus}} \right)^2
\left(\frac{\delta v}{\sigma_{\rm DH}}\right)^3, 
\ee
where $\delta{\bf v}=\delta v^3=\delta v_x \delta v_y \delta v_z\sim (0.5 {\rm km \ s^{-1}})^3$ is the 3D velocity bandwidth. 
For $\mathcal{A}\sim 10^{9}$, the amplified DM density $\rho_{\rm Earth}$ attains a value an order of magnitude greater than the non-amplified density $\rho_\odot$, and the amplified density has diurnal and seasonal modulations. 
It may be mentioned that everywhere on the Earth's surface is exposed at any time to such modulating focal flow of DM with $v_{\rm inj}=17$ \kms with the magnitude of amplification depending on the geographic longitude and latitude.

\subsection{Focal Length and Focusing Velocity}

The focal length of near-axis DM particles varies with the injection velocity, or the higher is the velocity, the more distant is the focus. Figure \ref{focallength} plots the focal length as a function of the injection velocity. 
The velocity of large-off axis particles that focus on the Earth's surface varies with the injection parameter. Figure \ref{focus-v-dr} plots the initial velocity of such DM as a function of the impact parameter.

We also mention that the DM flow is accelerated by the Earth's gravity, so that particles' velocity at the focus attains a slightly larger value than the injection velocity as $v_{\rm focus}=\sqrt{v_{\rm inj}^2+ 2 GM_{\rm E}/R_{\rm E}} $. For~example, the DM flow at $v_{\rm inj}= 17$ \kms focuses on the Earth's surface (detector) at \mbox{$v_{\rm focus}=20$ \kms.}

	\begin{figure}  
\begin{center}     
\includegraphics[width=8cm]{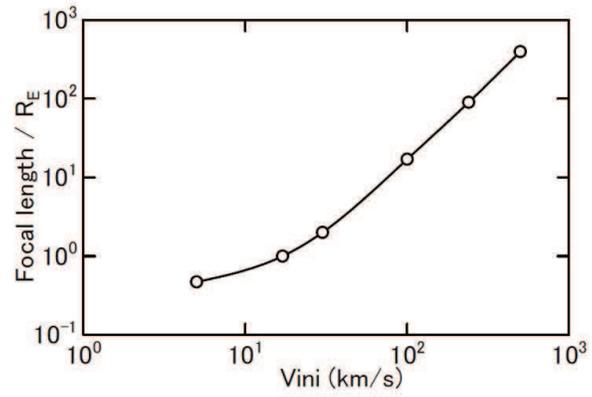} 
\end{center}
\caption{Focal length in unit of the Earth radius plotted against injection velocity for near-axis DM~orbits. }
   \label{focallength}
   \end{figure}

	\begin{figure} 
\begin{center}  
\includegraphics[width=7cm]{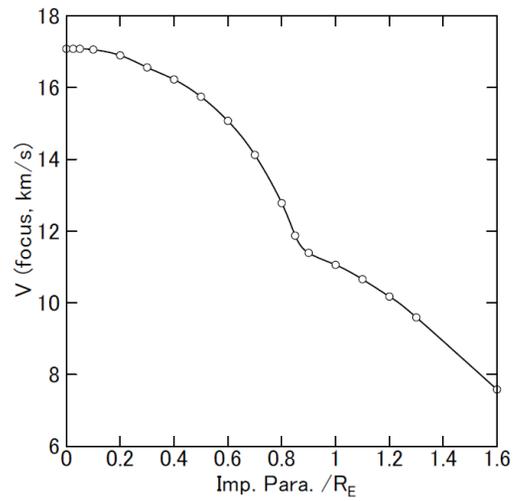} 
\end{center}
\caption{Injection velocity of DM focusing on the Earth's surface, plotted against impact~parameter. }     
\label{focus-v-dr}
   \end{figure}

\subsection{Other Focusing Objects}

The gravitational focusing applies to the other extended objects like the Moon, planets and the Sun \cite{Alenazi+2006,Hoffmann+2003,Patla+2014}, whose interior density distributions are well known . An extended object of radius $r_{\rm aperture}$ causes convex-lens like convergence onto a focal point with an amplification proportional to the area of the aperture (Eq. \ref{eqamp}). The angular radius, $\theta \sim r_{\rm aperture}/d$, of the aperture object at a distance $d$ is related to its mass and injection velocity through  
\be
\theta\sim \eta \left( \frac{G M}{d} \frac{1}{ v_{\rm inj}^2} \right)^{1/3} ,	 
\label{theta}
\ee
where $\eta\sim 0.8$ is a factor obtained by numerical calculation of orbits.
The Moon acts as a filled-aperture collector of DM flow with injection velocity $\sim 400$ \kms onto the Earth. The Sun and Jupiter are too massive, so that the injection velocity is required to be two order of magnitudes higher than $\sigma_{\rm DH}$. Extended galactic objects like molecular clouds are also possible filled-aperture collectors. The Orion molecular cloud ($d \sim 400$ pc, $r_{\rm aperture} \sim 15$ pc, $M \sim 10^5\Msun$), for example, can collect a DM flow of $\sim 180$ \kms with aperture diameter of $2\theta\sim 2^\circ$. 
Such apparent diameters may be compared to that of the Earth on the order of $\sim 30-60\deg$ as shown in the previous subsection. 

Alternatively, the objects may be used as concave-lens like collectors, with which a cylindrical DM flow makes an enhanced ring (small circle) around a point mass (\cite{Alenazi+2006}). 
The apparent radius is also given by Eq. \ref{theta} by replacing the angular radius by $\theta\sim r_{\rm ring}/d$. In case of the Sun ($d=1$ AU), a DM flow with injection velocity $\sim$200 \kms and radius (impact parameter) $\sim 60 R_\odot$ focuses on the Earth, making a DM ring of diameter $2\theta \sim 25^\circ$. 
The Moon, Jupiter, and the nearest star, $\alpha$ Cen, make rings of $2\theta \sim 0\deg.7$, $\sim 1\deg.6$ and $\sim 0\deg.4$, respectively, for $v_{\rm inj}\sim 200$ \kms.
The amplified DM density from the ring at the focal point (Earth) is on the order of
\be
\Delta \rho_{\rm ring} \sim 
\rho_\odot \left( \frac{r_{\rm ring}}{r_{\rm focus}} \right) 
\left(\frac{\delta v}{\sigma_{\rm DH}}\right)^3.
\ee  
We summarize the estimated values in table \ref{tab} of Appendix.

We finally mention that the parallactic aberration due to the relative orbital velocities between the Earth and the objects is large, which amounts to an order of $\sim$30\kms$/200$ \kms, causing significant displacement of the ring center toward the ecliptic east by $\sim 9^\circ$. For Jupiter, it is on the same order, but the aberration direction changes semi-annually.

\section{Summary } 
                
We considered possible local modulation of the DM density due to the Earth gravity. We showed that the flow of low-velocity DM at initial geo-centric velocity of $v_{\rm inj} = 17$ \kms focuses on the Earth's surface at the focusing velocity of 20 \kms where the flux is significantly amplified. 
Such amplification may have implication for direct detection experiments of DM and for the interpretation of the physics of detected DM flux. 

%%%%%%%%%%%%%%%%%%%%%%%%%%%%%%%%%%%%%%%%     
\vspace{6pt}
Funding: {This research received no external funding.}

%%%%%%%%%%%%%%%%%%%%%%%%%%%%%%%%%%%%%%%%%%
%\acknowledgments{ NONE in this work.  }
%%%%%%%%%%%%%%%%%%%%%%%%%%%%%%%%%%%%%%%%%%
Conflixt: {The author declares no conflict of interest. }  

%=====================================
% References, variant A: internal bibliography
%=====================================
%\reftitle{References}

\newpage
%%%%%%%%%%%%%%%%%%%%%%%%%%%%%%%%%%%%%%%%%%%%%
\begin{appendix}

\section*{Appendix: Table for estimated apertures and DM flow velocities}

Table \ref{tab} lists estimated injection velocities and apparent aperture/ring diameters for gravitational convergence onto the Earth by various objects.

\begin{table}  
\begin{center}
\caption{Injection velocity of DM for filled-aperture focusing on the Earth (surface), and angular diameter of focusing ring at injection velocity 200 \kms as seen from the Earth about various objects.} 
\label{tab} 
\begin{tabular}{ccccccc}  
\hline
\hline 
\textbf{Object} &  \textbf{Earth}& \textbf{Moon}&\textbf{Sun}&\textbf{Jupiter}&\textbf{$\alpha$ Cen}&\textbf{Ori. Mol. Cloud }\\

\hline
     Mass (g)& $5.97\times 10^{27}$ & $7.3\times 10^{25}$ & $1 \Msun$& $1.9\times 10^{30}$ &$0.91\Msun$& $\sim 10^{5} \Msun$\\
   
     Radius (km, pc)  & $6.37\times 10^3$ & $1.75\times 10^3$  & $6.9\times 10^5$ &$1.4\times 10^5$& $8.5\times 10^{5}$  & $\sim 15$ pc\\
     
     Distance (km, etc.)& $6.37\times 10^3$ & $3.8\times 10^5$& 1 AU &3.5--6.5 AU& 1.3 pc& $\sim 400$ pc\\
 
     \hline
     
     Fil. ap. dia. & $\sim 30\deg-60\deg(^\ddagger)$ & $0\deg.5$&$0\deg.5(^\dagger)$&$\sim 0\deg.01(^\dagger)$& ---& $\sim 2$\\
         
     Inj. velo. (\kms) & 17& $\sim 400$& $\gg \sigma_{\rm DH}$  & $\gg \sigma_{\rm DH}$ & ---& $\sim 180$ \\ 
   
   \hline
   
    Ring dia. for& --- & $0\deg.7$& $\sim 25\deg$&$\sim 1\deg.6$& $\sim 0\deg.4$& ---\\ 
    
    Inj. velo. (\kms)& --- & 200 & 200& 200& 200&---\\ 
\hline
\end{tabular} \\
$^\ddagger$ Figs. \ref{orbPandU}, \ref{focus}; $^\dagger$ \small{Uniform density is assumed, but actual msss in the Sun and Jupiter are more concentrated, so that effective apertures are smaller, and the velocities are higher. }
\end{center}
\end{table}
\unskip

\end{appendix}

   \end{document}